\documentclass[preprint,12pt]{elsarticle}

 \usepackage{graphicx}
 \usepackage{caption}
 \usepackage{subcaption}
 \usepackage[english]{babel}

\usepackage{amssymb}

\journal{Nuclear Physics A}

\begin{document}

\begin{frontmatter}



\title{X-ray beam-shaping via deformable mirrors: analytical computation of the required mirror profile}


\author[addr1]{D. Spiga}
\fntext[indir]{Phone : +39-039-5971027, Fax: +39-039-5971001}
\ead{daniele.spiga@brera.inaf.it}

\author[addr2]{L. Raimondi}
\author[addr2]{C. Svetina}
\author[addr2,addr3]{M. Zangrando}

\address[addr1]{INAF/Osservatorio Astronomico di Brera, Via E. Bianchi 46, I-23807, Merate - Italy}
\address[addr2]{Sincrotrone Trieste ScpA, S.S. 14 km 163.5 in Area Science Park, 34149 Trieste - Italy}
\address[addr3]{IOM-CNR, S.S. 14 km 163.5 in Area Science Park, 34149 Trieste - Italy}
	  
\begin{abstract}
X-ray mirrors with high focusing performances are in use in both mirror modules for X-ray telescopes and in synchrotron and FEL (Free Electron Laser) beamlines. A degradation of the focus sharpness arises in general from geometrical deformations and surface roughness, the former usually described by geometrical optics and the latter by physical optics. In general, technological developments are aimed at a very tight focusing, which requires the mirror profile to comply with the nominal shape as much as possible and to keep the roughness at a negligible level. However, a deliberate deformation of the mirror can be made to endow the focus with a desired size and distribution, via piezo actuators as done at the EIS-TIMEX beamline of FERMI@Elettra. The resulting profile can be characterized with a Long Trace Profilometer and correlated with the expected optical quality via a wavefront propagation code. However, if the roughness contribution can be neglected, the computation can be performed via a ray-tracing routine, and, under opportune assumptions, the focal spot profile (the Point Spread Function, PSF) can even be predicted analytically. The advantage of this approach is that the analytical relation can be reversed; i.e, from the desired PSF the required mirror profile can be computed easily, thereby avoiding the use of complex and time-consuming numerical codes. The method can also be suited in the case of spatially inhomogeneous beam intensities, as commonly experienced at Synchrotrons and FELs. In this work we expose the analytical method and the application to the beam shaping problem. 
\end{abstract}


\end{frontmatter}


\section{Introduction}
\label{sec:intro}
An important problem in X-ray optics is the prediction of the image quality from surface imperfections of a focusing, grazing-incidence mirror. This task is traditionally achieved via ray-tracing routines in the long spatial wavelengths spectral range, in which geometric optics is believed to be valid, and applying the known X-ray scattering theory in the high frequencies realm. Detailed information on the profile, and statistical information on the roughness of a grazing incidence mirror, enable the computation of the Point Spread Function (PSF) of the mirror \cite{RaiSpi 2010} or system of mirrors \cite{RaiSpi 2011} at any X-ray wavelength.

However, the inverse problem, i.e., the analysis of the PSF to derive the mirror profiles, has received less attention. While in the first-order scattering regime the theory allows an immediate inversion of the formalism, i.e., from the scattering distribution to the Power Spectral Density (PSD) of the surface roughness \cite{Church 1979}, in geometrical optics we still lack a direct, analytical method to connect the profile error to the shape of the PSF. Such a method could have application especially in {\it beam shaping} techniques that turn the radiation intensity distribution provided by a source like a FEL (Free Electron Laser) into another one on the focal plane, via a deformable mirror. Beam shaping techniques are being adopted at the EIS-TIMEX beamline of FERMI@Elettra \cite{Svetina 2011}.

\begin{figure}[!tpb]
        	\centering
         \includegraphics[width=0.8\textwidth]{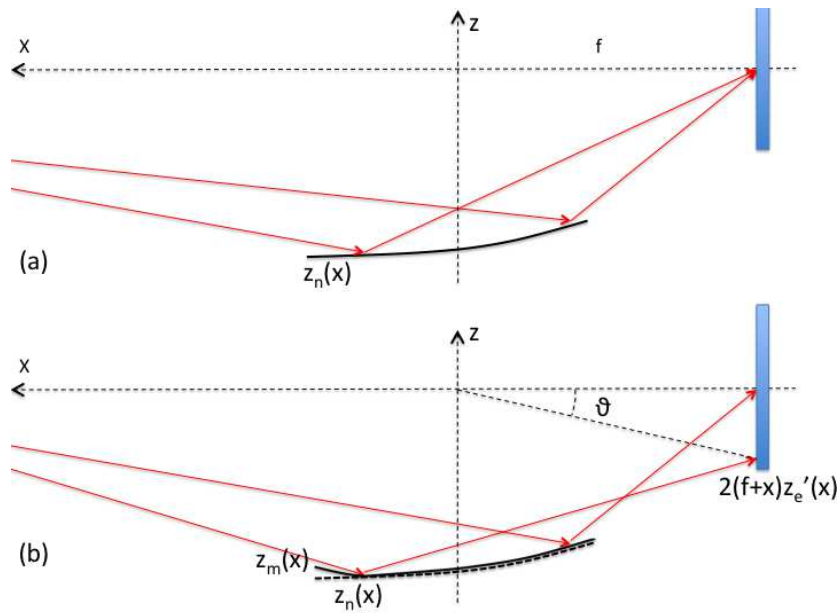}
         \caption{(a) A focusing mirror with ideal profile. (b) Profile errors spread the reflection over the focal plane.}
         \label{fig:fig01}
\end{figure}

In this paper we establish an analytical relation between the {\it simplest} profile of a single-reflection mirror characterized by a real longitudinal profile $z_{\mathrm m}(x)$ and its PSF in the {\it nominal} focal plane. We suppose the focal plane to be located at a distance $f$ from the mid-plane of the mirror (Fig.~\ref{fig:fig01}), assumed as origin of the reference frame, and we denote the nominal profile of the mirror with $z_{\mathrm n}(x)$, of length $L$ over the $x$-axis. The negative $x$ semi-axis points toward the focal plane. For on-ground X-ray sources like Synchrotrons or FELs, which are located at a large, but finite distance, the focusing profile is usually an ellipsoid. For astronomical sources, in practice at infinity, imaging is obtained via mirror systems including a paraboloid, usually followed by a second reflection on a hyperboloid \cite{VanSpey 1972}. In most cases, the nominal profile is designed to concentrate the beam to a single point on the focal plane, at (0, $-f$); the resulting PSF is a Dirac delta function. Sometimes, the mirror geometry deliberately allows some tolerable aberration on-axis, to improve the focus off-axis and so increase the mirror useful field \cite{Conconi 2001}. 

Indeed, real mirrors are characterized by profile defects that can be measured, e.g., using a Long Trace Profilometer \cite{Takacs 1999}. While the specific nominal shape affects the existence, the location, and the intrinsic aberrations of an {\it ideal} mirror focus, the PSF of a {\it real} mirror focus mostly depends on the mirror {\it profile error},
\begin{equation}
	z_{\mathrm e} = z_{\mathrm m}-z_{\mathrm n}.
	\label{eq:error}
\end{equation}
The resulting slope distribution spreads the PSF over the focal plane. The PSF also depends on the distribution of the intensity over the mirror length: astronomical sources illuminate uniformly the mirrors owing to their practically infinite distance.  Conversely, FEL sources are characterized by a markedly anisotropic brilliance, usually Gaussian. Therefore, the deformations of near the mirror center usually affect the PSF much more than those near the edges. 

A {\it ray-tracing routine} is the most widespread method adopted to compute the PSF from the X-ray mirror shape and illumination, in geometrical optics approximation, even if the results are difficult to express analytically. For simplicity, we heretofore suppose these 3 conditions to be fulfilled:
\begin{itemize}
	\item{the geometrical optics approximation is applicable, i.e., the effects of surface roughness and aperture diffraction are negligible;}
	\item{the roundness (sagittal) errors have a negligible impact on the PSF;}
	\item{the X-ray source can be assumed to be point-like.}
\end{itemize}
The first condition is not always easy to verify {\it a priori}, because the spectral range of spatial frequencies where the geometrical optics can be applied varies with the X-ray energy, the incidence angle $\alpha$, and even the amplitude of defects \cite{RaiSpi 2010, RaiSpi 2011, Aschen 2005}. 

The second condition is usually fulfilled in grazing incidence, because the roundness errors effect is suppressed by a factor of $\frac{1}{2}\tan2\alpha$ with respect to the axial errors. This allows us to reduce the relation profile-PSF to only one dimension, considering the sole, dominating effect of the axial profile errors, $z_{\mathrm e}(x)$. Also the PSF dependence is reduced to a single variable, the $\theta$ angle (Fig.~\ref{fig:fig01}).

The applicability of the third condition depends on the specific application and the accuracy requested in shaping the focused beam. In practice, it is met whenever the source, as imaged and demagnified on the focal plane, is much smaller than the desired beam size. 

\begin{figure}[!htpb]
	 \centering
	\begin{subfigure}{0.8\textwidth}     		 
       		   \includegraphics[width=\textwidth]{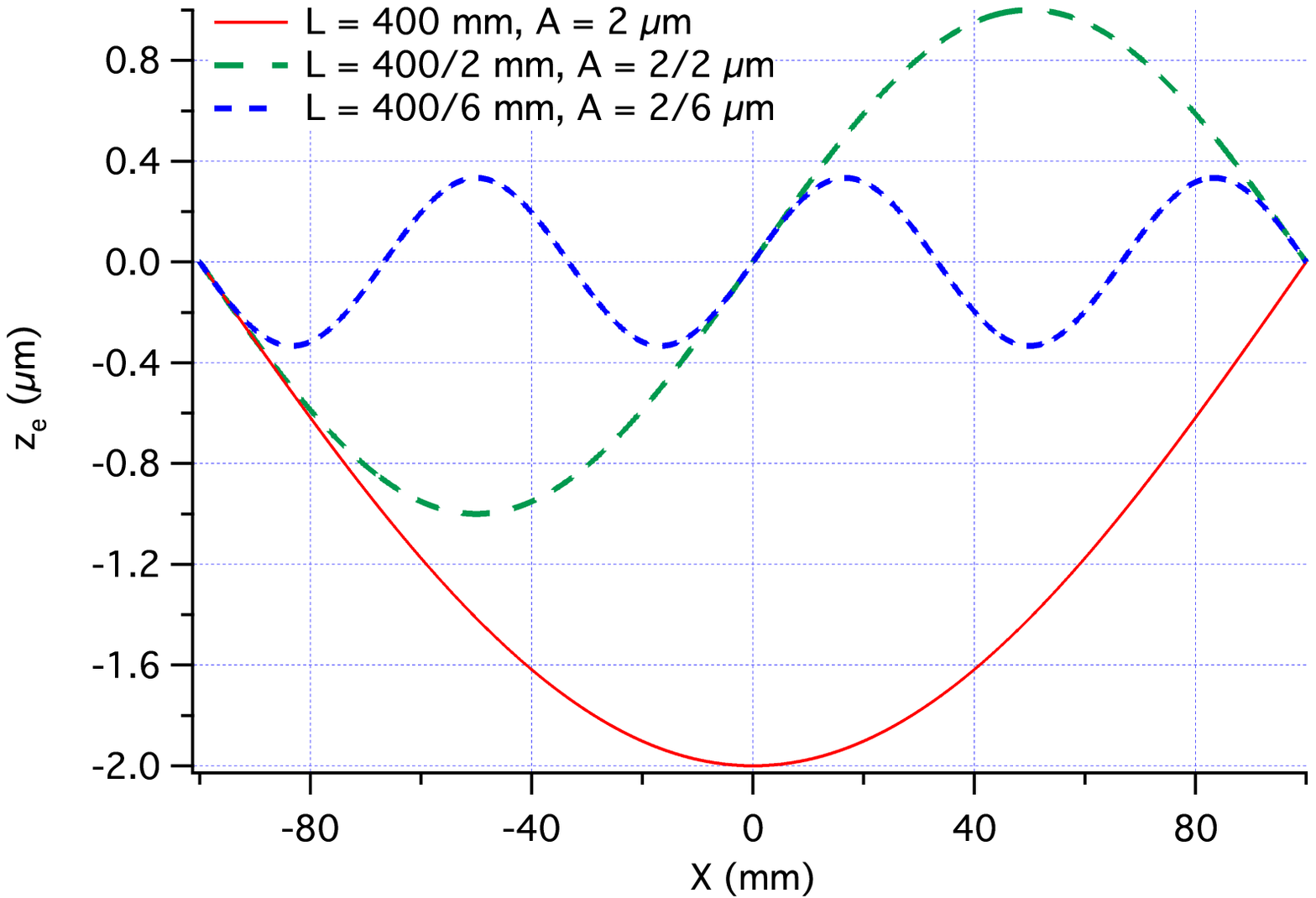}
     	            \caption{}
    	            \label{fig:fig02a}
         \end{subfigure}\\
         	\begin{subfigure}{0.8\textwidth}
         		\centering
         		\includegraphics[width=\textwidth]{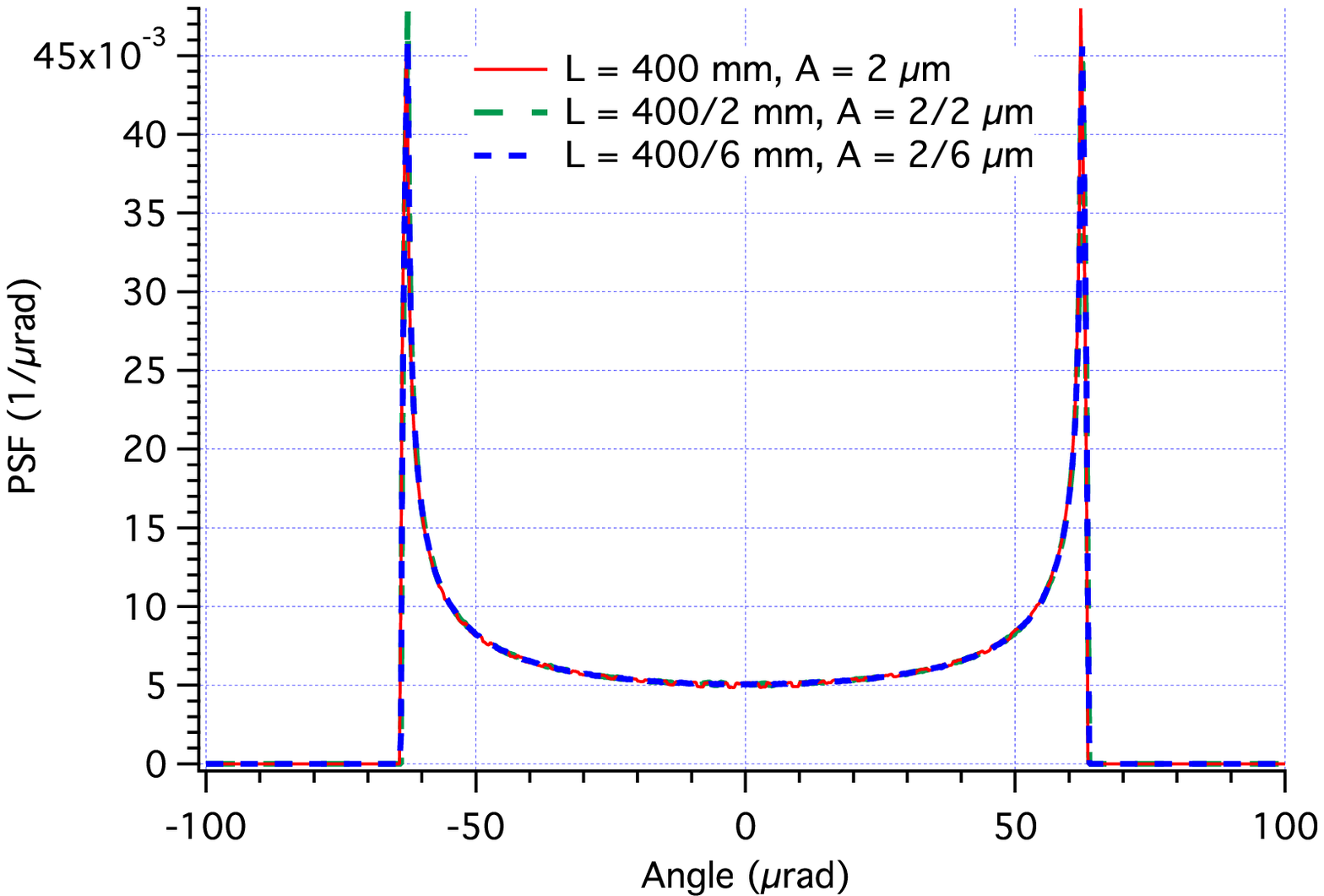}
         		\caption{}
      	   	\label{fig:fig02b}
         \end{subfigure}
         \caption{An example of PSF degeneracy: (a) sinusoidal profile errors $z_{\mathrm e}$ of different period, $2L$, and amplitude, $A$, but the same $A/L$ ratio, (b) return the same PSF (Eq.~\ref{eq:PSFsine}). Only the solid line, however, does not exhibit curvature inversions and is suitable for the calculation in Sect.~\ref{sec:analytical}. The beam intensity that impinges on the mirror is assumed to be uniform.}
         \label{fig:fig02}
\end{figure}

Although a PSF can be computed univocally via a ray-tracing routine,  there is in general more than one $z_{\mathrm e}(x)$ to return a single PSF (Fig.~\ref{fig:fig02}). The PSF results from angular deviations of rays on the surface, so it depends more on the slope distribution than on the mirror profile itself. Among all the possible mirror profiles, we thereby select the {\it only one} that fulfills the additional condition:
\begin{itemize}
\item{the derivative of the profile error, $z'_{\mathrm e}$, shall increase monotonically over the profile, i.e., the concavity of $z_{\mathrm e}$ must always be upwards.}
\end{itemize}
This condition is not related to intrinsic curvature of the mirror, whose nominal profile $z_{\mathrm n}(x)$ {\it must} be concave upwards in order to have focusing properties, but simply {\it operates a selection} among all the infinite possible profile errors, which in principle can have any concavity. For example, all the sinusoidal errors in Fig.~\ref{fig:fig02a} return the same PSF, but only the solid line fulfills the $z'_{\mathrm e}$ monotonicity condition. In the Sect.~\ref{sec:analytical} of this paper we see that this selection establishes a one-to-one correspondence between locations on the mirror and on the focal plane: this in turn allows us to set an analytical relation between a profile error with these properties and a PSF. Some examples of computation are shown in Sect.~\ref{sec:examples}. 

\section{Analytical relation between mirror profile and in-focus PSF}
\label{sec:analytical}
\subsection{From profile to PSF}\label{sec:direct}
We define $g(x)$ to be the beam intensity distribution over the mirror length, in the $x$ direction: we require that $g(x) \ge 0$ for all $-\frac{L}{2} < x < +\frac{L}{2}$ and that $g$ is normalized to 1 over the mirror length,
\begin{equation}
	\int_{-L/2}^{+L/2}\! g(x)\, \mbox{d}x = 1:
	\label{eq:normg}
\end{equation} 
if the beam distribution is uniform, $g(x) = 1/L$.

Rays impinging on the mirror at the coordinates $(x, z_{\mathrm m})$ are deviated from the focus according to the local slope. The angular deviation of the ray, owing to the small angle approximation, is twice the slope error, $z'_{\mathrm m}(x)-z'_{\mathrm n}(x) = z'_{\mathrm e}(x)$. After traveling a distance $f+x$ (Fig.~\ref{fig:fig01}), they intersect the focal plane with a lateral displacement
\begin{equation}
	z_{\mathrm F} = z_{\mathrm e}(x)+2(f+x)z'_{\mathrm e}(x);
	\label{eq:zfoc}
\end{equation}
as the deformations are usually below the micron and $f$ is of several metres, the 2$^{nd}$ term is overwhelmingly dominant. Hence, the angle seen from the centre of the optic (Fig.~\ref{fig:fig01}) is $\theta \simeq z_{\mathrm F}/f$, or
\begin{equation}
	\theta(x) \simeq 2\,\frac{f+x}{f}\,z'_{\mathrm e}(x).
	\label{eq:theta}
\end{equation}

\begin{figure}[!tbh]
        	\centering
         \includegraphics[width=0.75\textwidth]{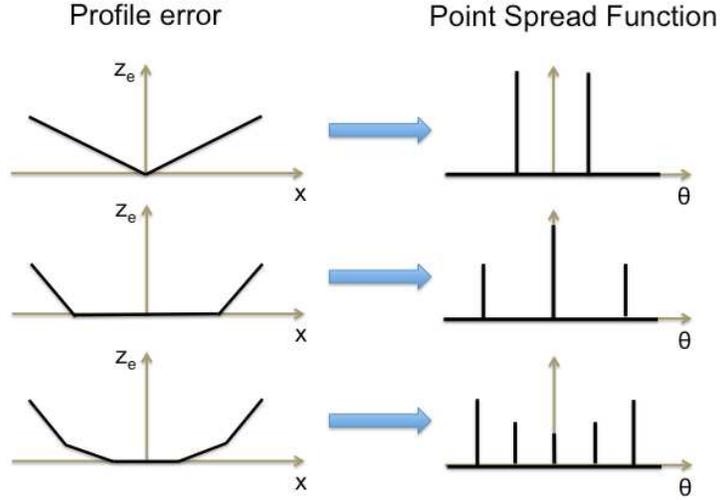}
         \caption{Profile errors with discrete steps in the derivative. Each step results in a delta-like peak in the PSF: the integral of each peak is proportional to the width of the respective step in the profile and to the beam intensity at that position. As the steps are reduced, the peaks become lower and closer, and tend to a continuous PSF.}
         \label{fig:fig03}
\end{figure}

By hypothesis, the derivative $z'_{\mathrm e}(x)$ is an increasing function of $x$, so $\theta$ is also an increasing function of $x$. Therefore, we have a one-to-one correspondence $x \leftrightarrow \theta$. In most cases, the mirror length is negligible with respect to the focal length, and Eq.~\ref{eq:theta} reduces to $\theta(x) \approx 2z'_{\mathrm e}(x)$.

We now divide the mirror profile into segments of variable length, $\Delta x_k$, with $k =0, 1, \cdots$ aiming at constant derivative changes, $\Delta z'_{\mathrm e}$ (Fig.~\ref{fig:fig03}). In this way, the derivative of the segments takes on discrete, equally-spaced values $z'_{{\mathrm e},k}$, increasing from its minimum $z'_{\mathrm e,m}$ at $x = -L/2$, to its maximum $z'_{\mathrm e,M}$ at $x = +L/2$. Likewise, $\theta$ increases as per Eq.~\ref{eq:theta} over the focal plane from its minimum, $\theta_{\mathrm m}$, to its maximum, $\theta_{\mathrm M}$. Rays striking on the mirror between $x_k$ and $x_k+\Delta x_k$ reach the focal plane at angles between $\theta_k$ and $\theta_k + \Delta\theta_k$, with an intensity $\Delta I(\theta_k)$ proportional to the length of the $k^{\mathrm{th}}$ segment and to $g(x_k)$,
\begin{equation}
	\Delta I(\theta_k) \propto g(x_k) \sqrt{(\Delta x_k)^2+(\Delta z_k)^2}\approx g(x_k)\,\Delta x_k.
	\label{eq:Idisc}
\end{equation}
The $\Delta\theta_k$'s are obtained by differentiating Eq.~\ref{eq:theta},
\begin{equation}
	\Delta\theta_k = 2\,\frac{f+x_k}{f}\,\Delta z'_{\mathrm e} + \frac{2}{f}\,\Delta z_{{\mathrm e},k}:
	\label{eq:deltatheta}
\end{equation} 
owing to the large focal length, the second term of Eq.~\ref{eq:deltatheta} is negligible. Clearly, $\Delta\theta_k > 0$. The PSF is the intensity in the nominal focal plane per angle unit, $\Delta I(\theta_k) /\Delta\theta_k$:
\begin{equation}
	PSF(\theta_k) = g(x_k) \frac{\Delta x_k}{\Delta \theta_k} = g(x_k)\frac{f}{2(f+x_k)}\left(\frac{\Delta z'_{{\mathrm e}}}{\Delta x_k}\right)^{-1}.
	\label{eq:Pdisc}
\end{equation}
Passing to the actual profile in the limit $\Delta z'_{\mathrm e} \rightarrow 0$, Eq.~\ref{eq:Pdisc} turns into
\begin{equation}
	PSF(\theta) = \left.\frac{g_{\mathrm a}(x)}{2\,z_{\mathrm e}''(x)}\right|_{x = x(\theta)},
	\label{eq:dirform}
\end{equation}
for all values of $\theta$ that correspond to some $x$ via Eq.~\ref{eq:theta} -- that can be inverted because of the supposed $z'_{\mathrm e}(x)$ monotonicity -- and zero elsewhere. To simplify the notation, in Eq.~\ref{eq:dirform} we have introduced the {\it asymmetric} intensity distribution,
\begin{equation}
	g_{\mathrm a}(x) = \frac{f}{f+x}\,g(x).
	\label{eq:asymm}
\end{equation}
In the limit of a regular profile, Eq.~\ref{eq:deltatheta} becomes
\begin{equation}
	\theta'(x) = 2\,\frac{f+x}{f}\, z''_{\mathrm e}(x);
	\label{eq:thetaderi}
\end{equation}
therefore Eq.~\ref{eq:dirform}, as expected, is normalized to 1 over the focal plane:
\begin{equation}
	\int_{\theta_{\mathrm m}}^{\theta_{\mathrm M}} \!PSF(\theta)\, \mbox{d}\theta = \int_{z'_{\mathrm e,m}}^{z'_{\mathrm e,M}}\! g(x)\, \frac{\mbox{d}z'_{\mathrm e}}{z''_{\mathrm e}} = \int_{-L/2}^{+L/2}\!g(x) \,\mbox{d}x =1.
	\label{eq:Pnorm}
\end{equation}
As $z'_{\mathrm e}(x)$ is an increasing function of $x$, $z''_{\mathrm e}(x) \ge 0$. Hence, Eq.~\ref{eq:dirform} returns a positive and finite PSF wherever $z''_{\mathrm e}(x) > 0$. If $z''_{\mathrm e}(x) = 0$ at some $x$, however, some attention should be paid when applying Eq.~\ref{eq:dirform}:
\begin{itemize}
	\item{If $z''_{\mathrm e}(x) = 0$ at some {\it isolated} $x = \bar{x}$, i.e., the profile error tend to be straight as $\bar{x}$ is approached, then the PSF diverges to infinity at $\theta(\bar{x})$ but Eq.~\ref{eq:dirform} can be applied at all the nearby locations. In practice, this produces a sharp peak, but not a Dirac delta, in the PSF. Anyway, the integral of the PSF remains finite (Eq.~\ref{eq:Pnorm}). }
	\item{If $z''_{\mathrm e}(x) = 0$ in one or more {\it intervals} of the kind [$x_j$, $x_j+\Delta x_j$], with $j = 1, 2,\ldots, N$, then $z_{\mathrm e}(x)$ exhibits $N$ straight segments and $\theta(x)$ has $N$ ''plateaux'', i.e., $\theta = \theta(x_j)$ in correspondence of those positions. Hence, the relation $\theta = \theta(x)$ cannot be reversed anymore. Physically, all the intensity impinging on the linear segments is focused at $\theta(x_j)$, resulting in Dirac deltas at those locations. To extend the computation to this case, we have two possibilities:
	\begin{enumerate}
		\item{$g(x) = 0$ wherever $z''_{\mathrm e}(x) = 0$. Therefore, the straight segments do not give contribution to the PSF and Eq.~\ref{eq:dirform} can be used: fortunately, profile deformations obtained from the inverse formula (Eq.~\ref{eq:invform}) fall precisely in this case.} 
		\item{$g(x) > 0$ in some interval where $z''_{\mathrm e}(x) = 0$. In this case, the straight segments shall be firstly removed from the profile before applying Eq.~\ref{eq:dirform}. The contribution of the straight segments, $PSF_{\mathrm{str}}$, should be subsequently included by adding $N$ Dirac deltas to the PSF, 
		\begin{equation}
			PSF_{\mathrm{str}}(\theta) = \sum_{j =1}^N \delta(\theta-\theta(x_j)), 
			\label{eq:PSFlin}
		\end{equation}
		each of them normalized to the total intensity impinging on the $j$-th segment:
		\begin{equation}
			\int_{\theta_{\mathrm m}}^{\theta_{\mathrm M}} \! \delta(\theta-\theta(x_j)) \,\mbox{d}\theta = \int_{x_j}^{x_j+\Delta x_j}\!\!\!\!\!g(x) \,\mbox{d}x.
			\label{eq:PSFlin_norm}
		\end{equation}}
	\end{enumerate}}
	\item{If $z_{\mathrm e}'(x)$ is {\it discontinuous} at some $x=\bar{x}$ (a ''kink'' in the profile) like in Fig.~\ref{fig:fig04b}, then $z_{\mathrm e}''(x)$ is infinite at $\bar{x}$, so $PSF(\theta) =0$ for all values of $\theta$ in the interval corresponding via Eq.~\ref{eq:theta} to the $z_{\mathrm e}'(x)$ leap at $\bar{x}$.}
\end{itemize}

\subsection{From PSF to profile}\label{sec:inverse}
\begin{figure}[!tpb]
	\centering
	\begin{subfigure}{0.8\textwidth}
	         	  \centering
       		   \includegraphics[width=\textwidth]{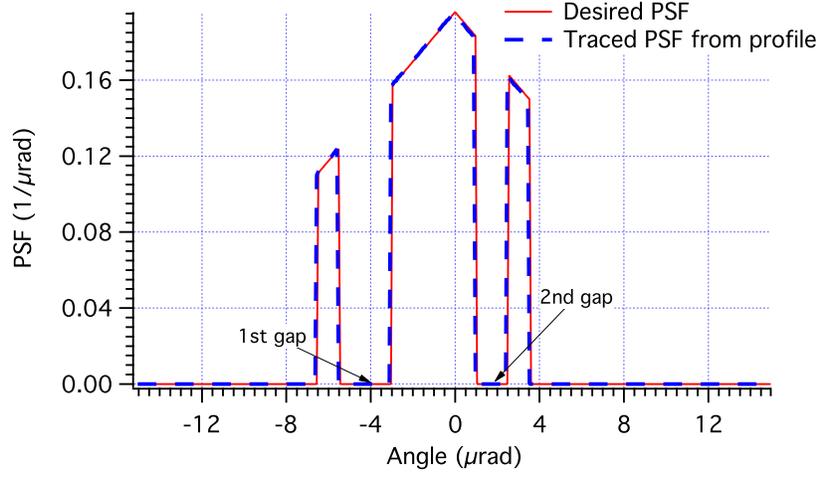}
     	            \caption{}
    	            \label{fig:fig04a}
         \end{subfigure}\\
         	\begin{subfigure}{0.8\textwidth}
         		\centering
         		\includegraphics[width=\textwidth]{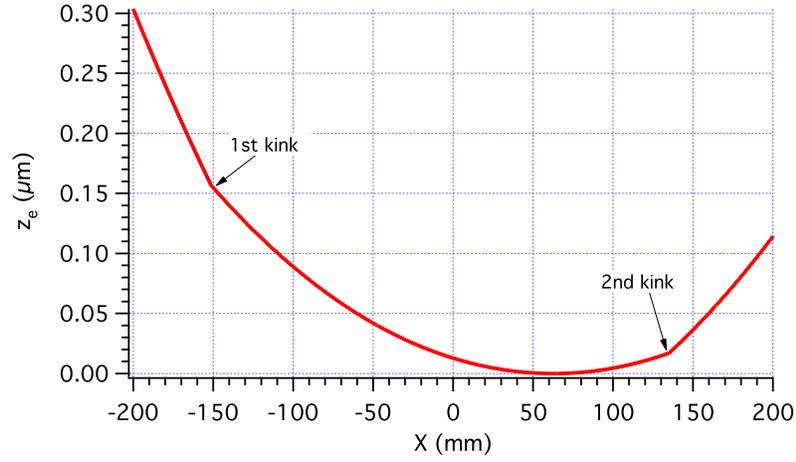}
         		\caption{}
      	   	\label{fig:fig04b}
         \end{subfigure}
         \caption{(a) An unusual PSF (solid line) with two gaps. (b) The profile deformation imparted to the mirror is computed from the requested PSF, via Eq.~\ref{eq:invform}, assuming an isotropic X-ray source, and $f \gg L$. The profile exhibits two discontinuities in $z'_{\mathrm e}$ ("kinks") corresponding to the two gaps in the PSF. Applying a ray-tracing routine to the computed profile yields back the initial PSF shown in Fig.~\ref{fig:fig04a} (dashed line).}
         \label{fig:fig04}
\end{figure}
We now derive the reverse form of Eq.~\ref{eq:dirform}, which can be directly applied to the problem of beam shaping (Sect.~\ref{sec:invcomp}). For any $PSF(\theta) \ge 0$ at the focal plane defined at angles [$\theta_{\mathrm m}, \theta_{\mathrm M}$], and for any $g(x) \ge 0$ over the mirror length [$-L/2$, $+L/2$], a profile error can be computed under the conditions listed in Sect.~\ref{sec:intro}. The substitution of Eq.~\ref{eq:asymm} and~\ref{eq:thetaderi} reduces Eq.~\ref{eq:dirform} to
\begin{equation}
	PSF(\theta)\cdot\theta'(x) = g(x).
	\label{eq:inv1}
\end{equation}
This simple differential equation is solved if $z'_{\mathrm e}(x)$ is known at one value of $x$. This choice affects the average tilt of the profile, hence it will change the position, but not the shape of the PSF: for instance, a natural choice for a symmetric PSF could be $z'_{\mathrm e}(0) = 0$. However, since both distributions are normalized to 1, it is convenient to set $z'_{\mathrm e}\left(-\frac{L}{2}\right) = \theta_{\mathrm m}$. This choice preserves also the position of an asymmetric PSF on the angular scale. 

Hence, integrating both sides of Eq.~\ref{eq:inv1}, we have 
\begin{equation}
	\int_{\theta_{\mathrm m}}^{2\frac{f+x}{f}z'_{\mathrm e}} \!\!\!\!\!PSF(\theta)\,\mbox{d}\theta = \int_{-\frac{L}{2}}^x \!g(t)\,\mbox{d}t,
	\label{eq:invform}
\end{equation}
and, if the integrations can be performed, the equation can be solved for $z_{\mathrm e}'(x)$ for any $x$ in the interval $[-L/2, +L/2]$. Finally, a further integration returns $z_{\mathrm e}(x)$. For Eq.~\ref{eq:invform} to be applied, the PSF must be normalized to 1 over the focal plane, and $g(x)$ must be normalized to 1 {\it over the mirror length} (Eq.~\ref{eq:normg}), not over the entire $x$-axis, otherwise the profile error will return only the core of the PSF: i.e., the part of the PSF that is normalized to the integral of $g(x)$ over the mirror length.

As long as both $PSF(\theta) > 0$ for all $\theta$ and $g(x) > 0$ for all $x$, the integrals in Eq.~\ref{eq:invform} are increasing functions of the upper integration limits. Therefore, solving for $z_{\mathrm e}'$ will automatically return an increasing function of $x$, i.e., an upwards concave profile error as required (Sect.~\ref{sec:intro}). Likewise the direct computation (Sect.~\ref{sec:direct}), a few special cases deserve some attention:
\begin{itemize}
 	\item{$g(x) = 0$ for some $x$: this occurs, e.g., if the beam extension over $x$ is smaller than the mirror length. In this case, solving Eq.~\ref{eq:invform}, $z'_{\mathrm e}(x)$ will be a constant where $g(x) = 0$, resulting in straight segments of $z_{\mathrm e}$. This appears reasonable, because there is nothing to shape where the intensity is zero. As we anticipated in the previous section, in the straight segments $z''_{\mathrm e}(x) = 0$, but also $g(x) = 0$, so the re-application of Eq.~\ref{eq:dirform} to the generated profile does not pose a problem.} 
	\item{$PSF(\theta) = 0$ at some $\theta$, like in the example shown in Fig.~\ref{fig:fig04a}: this situation leads to inversion problems, because the left-hand of Eq.~\ref{eq:invform} becomes a constant in an interval of $z'_{\mathrm e}$ for a single $\bar{x}$. A way to extend the computation to this case is to adopt for $z'_{\mathrm e}(\bar{x})$ the limit value {\it at the closest edge} of the intervals where left-hand of Eq.~\ref{eq:invform} is constant. In most cases, these intervals are in the PSF wings and do not pose a problem. Otherwise, they result in a $z'_{\mathrm e}$ discontinuity at $\bar{x}$, i.e. a kink in $z_{\mathrm e}$ (Fig.~\ref{fig:fig04b}).}
\end{itemize}

\section{Some examples}
\label{sec:examples}
\subsection{A direct computation}\label{sec:dircomp}
We hereafter consider some application of Eq.~\ref{eq:invform}, assuming for simplicity $f \gg L$, i.e., which implies the asymmetry factor (Eq.~\ref{eq:asymm}) to be completely negligible. For example, if the profile error is a half-sinusoid
\begin{equation}
	z_{\mathrm e}(x) = -A\cos\left(\frac{\pi}{L}x\right),
	\label{eq:sine}
\end{equation}
and the beam is initially uniform, Eq.~\ref{eq:dirform} returns a typical, diverging PSF:
\begin{equation}
	PSF(\theta) = \frac{1}{\pi}\left[\left(\frac{2\pi A}{L}\right)^2\!\!-\theta^2\right]^{-1/2},
	\label{eq:PSFsine}
\end{equation}
with $|\theta| < 2\pi A/L$, and zero elsewhere. A PSF of this kind was shown in Fig.~\ref{fig:fig02b}. The HEW  (Half Energy Width, i.e. the angular diameter including half integral of the PSF) can be directly computed from Eq.~\ref{eq:PSFsine}:
\begin{equation}
	HEW =  \sqrt{8}\pi \frac{A}{L}.
	\label{eq:HEWsine}
\end{equation}

\subsection{Some inverse computations}\label{sec:invcomp}
\begin{figure}[!tpb]
          \centering
	\begin{subfigure}{0.8\textwidth}
	         	  \centering
       		   \includegraphics[width=\textwidth]{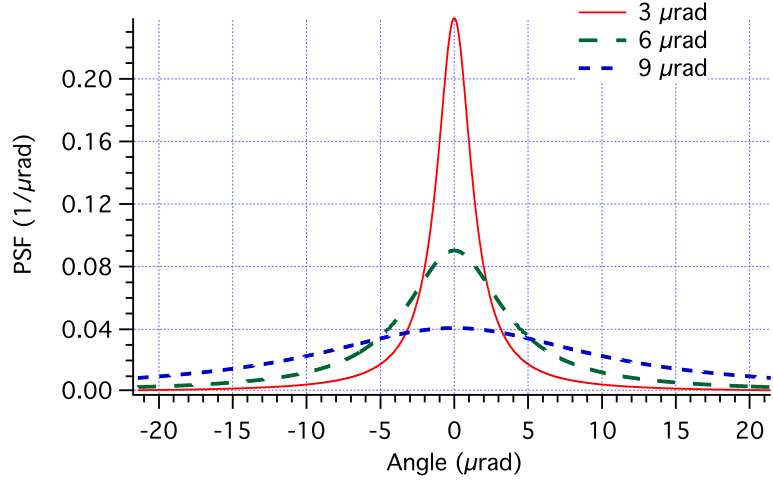}
     	            \caption{}
    	            \label{fig:fig05a}
         \end{subfigure}\\
         	\begin{subfigure}{0.8\textwidth}
         		\centering
         		\includegraphics[width=\textwidth]{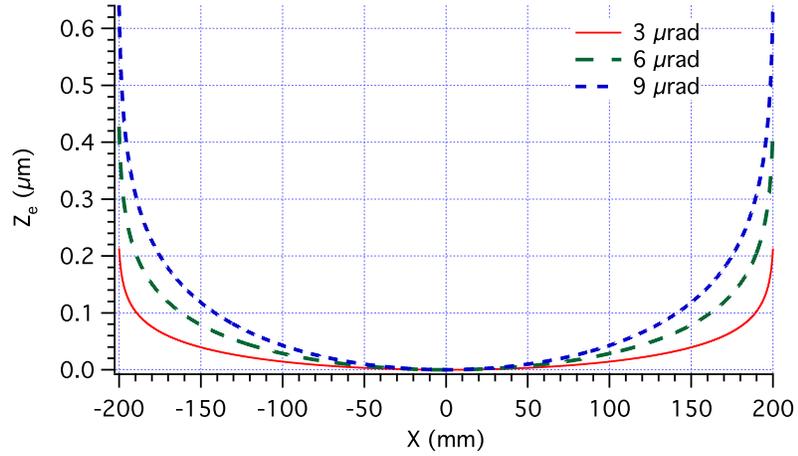}
         		\caption{}
      	   	\label{fig:fig05b}
         \end{subfigure}
         \caption{(a) Some Lorentzian PSFs (Eq.~\ref{eq:PSFlor}) characterized by different widths. (b) The deformations to be superposed to the nominal profile (Eq.~\ref{eq:zprof1}) to obtain the Lorentzian PSF, if the mirror is uniformly illuminated and $f \gg L$.}
         \label{fig:fig05}
\end{figure}
Inverse computations are more interesting, because they can be directly applied to beam-shaping problems. For instance, if a uniform beam impinges on a focusing mirror, which deformation yields a Lorentzian-shaped PSF,
\begin{equation}
	PSF(\theta) = \frac{2w}{\pi(w^2+4\theta^2)},
	\label{eq:PSFlor}
\end{equation}
with $w$ -- the HEW parameter -- arbitrary? For simplicity, we have assumed the focal plane to be infinitely extended, so Eq.~\ref{eq:PSFlor} is normalized to 1. The substitution of $g(x) = 1/L$ and Eq.~\ref{eq:PSFlor} into Eq.~\ref{eq:invform} yields, neglecting the asymmetry factor (Eq.~\ref{eq:asymm}) and assuming $z'_{\mathrm e}(0) = 0$ because of the PSF symmetry, 
\begin{equation}
	\arctan\left(\frac{4z'_{\mathrm e}}{w}\right) = \frac{\pi x}{L}.
	\label{eq:PSFlorint}
\end{equation}
Then, Eq.~\ref{eq:PSFlorint} can be solved for $z'_{\mathrm e}$:
\begin{equation}
	z'_{\mathrm e}(x) =\frac{w}{4}\tan\left(\frac{\pi x}{L}\right),
	\label{eq:zderi1}
\end{equation}
which returns by integration the desired profile
\begin{equation}
	z_{\mathrm e}(x) = -\frac{Lw}{4\pi}\log\cos\left(\frac{\gamma x}{L}\right),
	\label{eq:zprof1}
\end{equation}
with $\gamma \lesssim \pi$ to avoid the profile divergence at the mirror edges. The exact choice of $\gamma$ really affects only the very edges of the mirror profile: a suitable value can be e.g., $\gamma$ = 3.12. In Fig.~\ref{fig:fig05a} we display some Lorentzian PSFs for different values of $w$, and the respective profile errors are reported in Fig.~\ref{fig:fig05b}.

\begin{figure}[!htpb]
        	\centering
	\begin{subfigure}{0.8\textwidth}
	          \includegraphics[width=\textwidth]{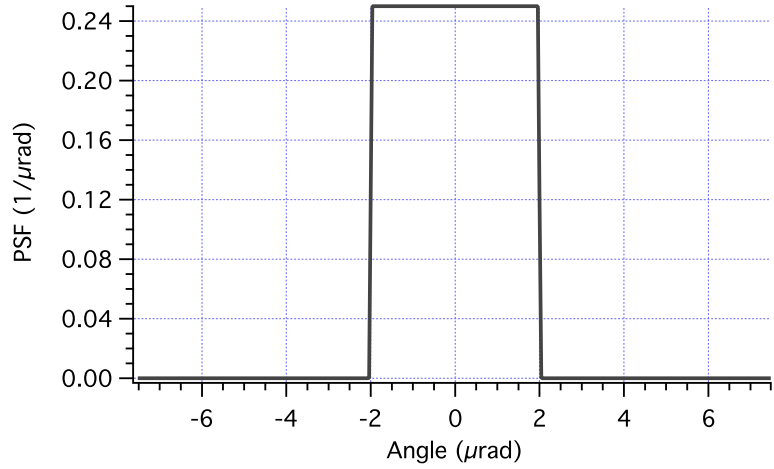}
     	          \caption{}
    	          \label{fig:fig06a}
         \end{subfigure}\\
	\begin{subfigure}{0.8\textwidth}
         		\includegraphics[width=\textwidth]{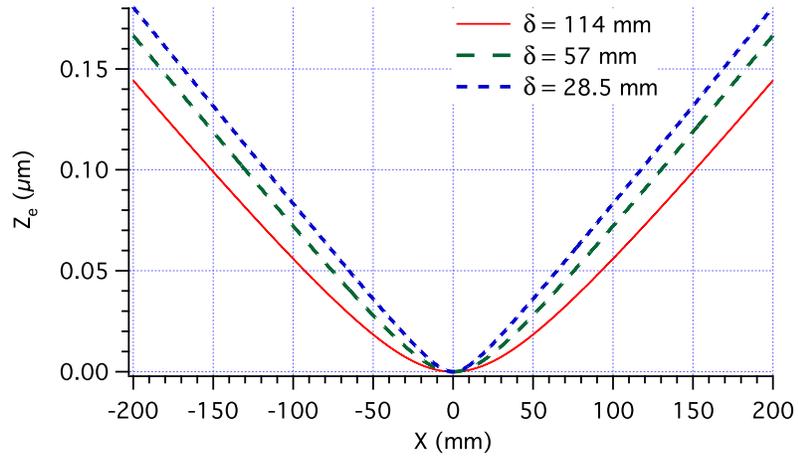}
                   \caption{}
    	          \label{fig:fig06b}
         \end{subfigure}
         \caption{(a) A top-hat distribution of 4 $\mu$rad full width (Eq.~\ref{eq:flattop}) and (b) the profile deformations (Eq.~\ref{eq:zprof2}) to turn a Lorentzian $g(x)$ of variable width $\delta$ into that PSF. We note that the profiles tend to become straight as $g(x) \rightarrow 0$ toward the edges of the mirror profile.}
         \label{fig:fig06}
\end{figure}

We now consider an initially anisotropic source (like a FEL)  that has to be turned into, e.g., a top-hat PSF of full width $w$ (Fig.~\ref{fig:fig06a}):
\begin{equation}
	PSF(\theta) = \frac{1}{w} \hspace{1cm}\mbox{for}\hspace{1cm}-\frac{w}{2}<\theta<+\frac{w}{2}.
	\label{eq:flattop}
\end{equation}
For simplicity, $g(x)$ is supposed to be a Lorentzian function over the mirror length,
\begin{equation}
	g(x) = \frac{2\delta}{\pi(\delta^2+4x^2)}.
	\label{eq:Glor}
\end{equation}
where $\delta$ is the HEW parameter. If $\delta \ll L$, $g(x)$ is nearly normalized to 1 over the mirror length. Substituting Eqs.~\ref{eq:flattop} and~\ref{eq:Glor} into Eq.~\ref{eq:invform}, we derive, neglecting the asymmetry factor (Eq.~\ref{eq:asymm}),
\begin{equation}
	z'_{\mathrm e}(x) =\frac{w}{2\pi}\arctan\left(\frac{2 x}{\delta}\right),
	\label{eq:zderi2}
\end{equation}
where we have set, owing to the PSF symmetry, $z'_{\mathrm e}(0)=0$. Eq.~\ref{eq:zderi2} can be easily integrated by parts, yielding
\begin{equation}
	z_{\mathrm e}(x) = \frac{w}{2\pi}\left[x\arctan\left(\frac{2 x}{\delta}\right)-\frac{\delta}{4}\log\left(1+\frac{4x^2}{\delta^2}\right)\right]:
	\label{eq:zprof2}
\end{equation}
some profiles of this kind are shown in Fig.~\ref{fig:fig06b}. The beam-shaping problem under study at the EIS-TIMEX beamline of FERMI \cite{Svetina 2011} requires the same computation, even if the initial intensity matches more a Gaussian than a Lorentz function. Since a Gaussian function cannot be integrated analytically, the profile computation has to be performed by solving Eq.~\ref{eq:invform} numerically. 

\begin{figure}[!htpb]
        	\centering
         	\begin{subfigure}{0.8\textwidth}
	          \includegraphics[width=\textwidth]{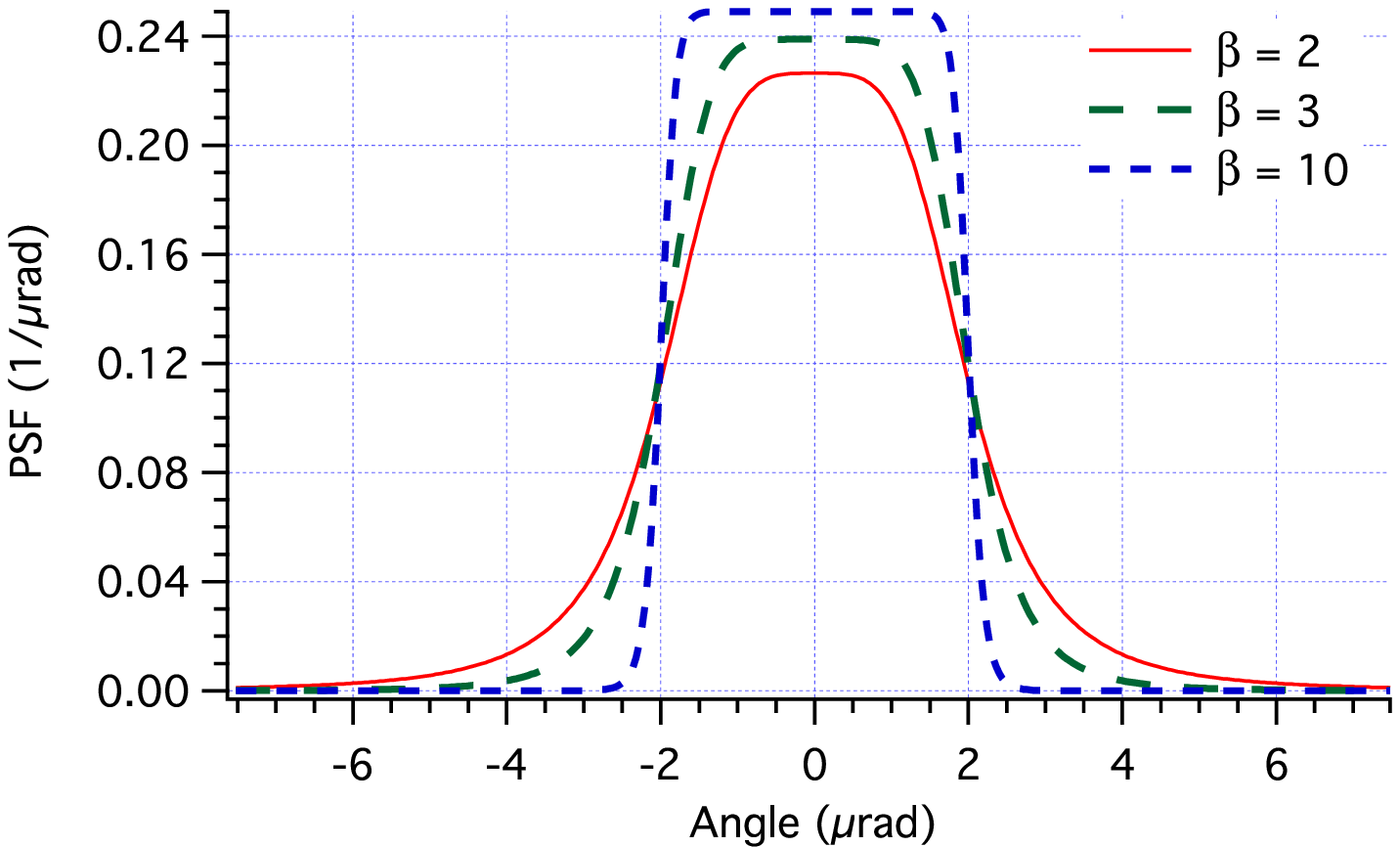}
     	          \caption{}
    	          \label{fig:fig07a}
         \end{subfigure}\\
	\begin{subfigure}{0.8\textwidth}
         		\includegraphics[width=\textwidth]{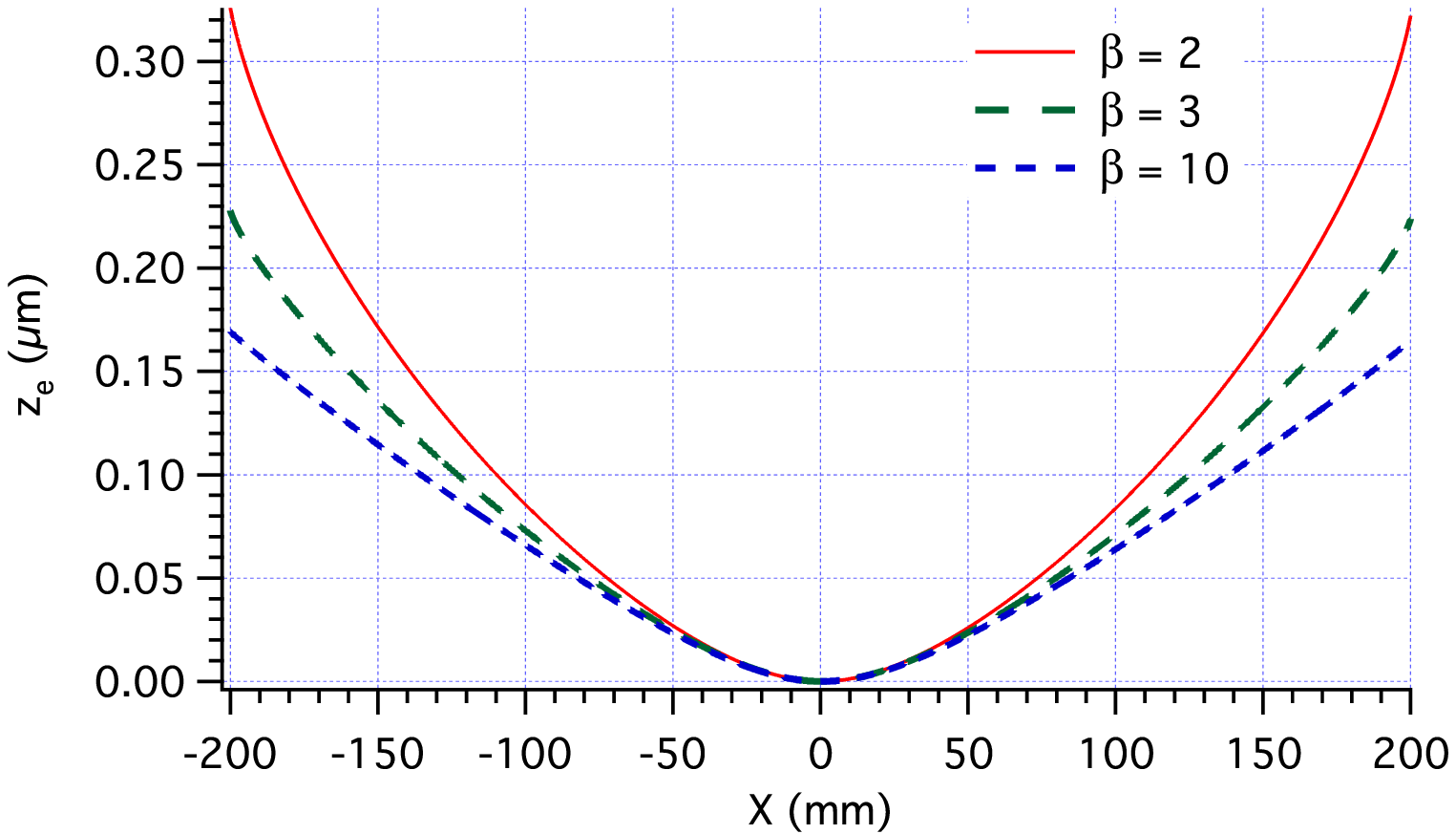}
                   \caption{}
    	          \label{fig:fig07b}
         \end{subfigure}
         \caption{(a) King-shaped PSFs (Eq.~\ref{eq:king}) with $w$ = 4 $\mu$rad for different values of the $\beta$ parameter. (b) The required profile deformations to obtain the PSFs if $g(x)$ is a Lorentzian (Eq.~\ref{eq:Glor}) with $\delta$ = 57 mm.}
         \label{fig:fig07}
\end{figure}

In Fig.~\ref{fig:fig07} we show the same exercise aiming at a PSF with King profile, i.e., a generalization of Eq.~\ref{eq:Glor}:
\begin{equation}
	PSF(\theta) = \frac{2N}{\pi w[1+(2\theta/w)^{2\beta}]},
	\label{eq:king}
\end{equation}
where $w$ is a width parameter, $\beta > 0$ is a slope parameter and $N$ is a normalization factor over the focal plane. We suppose $g(x)$ to still be a Lorentzian: the integration of a King function cannot be performed analytically, so the computation of the mirror deformation has to be obtained numerically. The PSFs and the corresponding mirror profiles are shown in Fig.~\ref{fig:fig07}. We note that the mirror profiles tend to resemble the ones of Fig.~\ref{fig:fig06} as the PSF converges to a top-hat shape, i.e., for increasing values of $\beta$.

\section{Conclusions}
In this work we have shown how the well-known problem of computing a focusing mirror PSF from the profile errors can be solved analytically for the special class of upwards-concave profiles. Although this seems less useful than a ray-tracing, a great advantage is the possibility to invert the formalism and derive a mirror profile deformation from a desired PSF. This can be done analytically in a number of cases, and numerically in any case. The resulting profile is the simplest possible one, as it does not exhibit undulations, therefore it should be possible to reproduce, e.g., via piezo actuators. Only if a PSF exhibits gaps, the profile will present kinks that can pose more difficulties to the shaping. It should be kept in mind that the results may differ from predictions if the initial hypotheses are not fulfilled: 
\begin{itemize}
	\item{If the geometrical optics is not completely applicable, a self-consistent treatment that includes the effect of aperture, mid-frequency defects, and roughness \cite{RaiSpi 2010, RaiSpi 2011} has to be considered, but the formalism becomes very difficult to reverse.}
	\item{If the sagittal errors are not negligible, the PSF has to be computed via a complete ray-tracing routine, which is also difficult to reverse.}
	\item{If the source size is not negligible, the demagnified source has to be convolved with the ideal PSF. Therefore, to compute the correct profile deformation from a requested PSF, the demagnified source profile has to be preliminarily de-convolved, which is not always possible.}
\end{itemize}
The extension of the formalism hitherto presented to the mentioned situations, and the computation for the case of EIS-TIMEX, will be the subject of a subsequent paper.





\bibliographystyle{model1-num-names}
\bibliography{<your-bib-database>}

\begin{thebibliography}{100}
\bibitem{RaiSpi 2010}
	L. Raimondi, D. Spiga, ``Self-consistent computation of x-ray mirror point spread functions from surface profile and roughnessÓ, Proc. SPIE. Vol. 7732, 77322Q (2010) 
\bibitem{RaiSpi 2011}
	L. Raimondi, D. Spiga, ``Point Spread Function of real Wolter-I X-ray mirrors: computation by means of the Huygens-Fresnel principleÓ, Proc. SPIE. Vol. 8147, 81470Z (2011)
\bibitem{Church 1979}
	E. L. Church, H. A. Jenkinson, J. M. Zavada, ``Relationship between surface scattering and microtopographic features'', Opt.~Eng., 18, 125-136 (1979)
\bibitem{Svetina 2011}
	C. Svetina, G. Sostero, R. Sergo, et al., ``A beam-shaping system for TIMEX beamlineÓ, NIM-A, Vol. 635 (1) S12-S15 (2011)
\bibitem{VanSpey 1972}
	L. P. Van Speybroeck, R. C. Chase, ``Design parameters of paraboloid-hyperboloid telescopes for X-ray astronomy'', Appl.~Opt., 11, 2, 440-445 (1972)
\bibitem{Conconi 2001}
	P. Conconi, S. Campana, ``Optimization of grazing incidence mirrors and its application to surveying X-ray telescopes'', A\& A, 372, 1088-1094 (2001)
\bibitem{Takacs 1999}
	 P. Z. Tak\'{a}cs, S. Qian, T. Kester, et al., ``Large-mirror figure measurement by optical profilometry techniques''. Proc. SPIE, Vol. 3782, p. 266-274 (1999)
\bibitem{Aschen 2005}
	B. Aschenbach, ``Boundary between geometric and wave optical treatment of x-ray mirrors''. Proc. SPIE, Vol 5900, p. 59000D1-7 (2005)
\end{thebibliography}



\end{document}